\documentclass{article}
\usepackage[utf8]{inputenc}

\title{Utilizing Expert Opinion to inform Extrapolation of Survival Models}
\author{Philip Cooney \& Arthur White}

\date{%
    School of Computer Science and Statistics, Trinity College Dublin\\%
    \today
}

\usepackage{pbox}
\usepackage{natbib}
\usepackage{graphicx}
\usepackage{amsmath}
\usepackage{float}
\usepackage{bm}
\usepackage{appendix}
\usepackage[margin=1in]{geometry}
\usepackage[table]{xcolor}
\usepackage{amssymb,amsmath}
\usepackage{url}
\usepackage{tabularx}
\usepackage{pdflscape}
\usepackage{afterpage}
\usepackage{capt-of}
\usepackage{fourier} 
\usepackage{array}
\usepackage{makecell}
\usepackage{hyperref}
\usepackage{multirow}
\usepackage{pdflscape}
\usepackage{subcaption}
\usepackage{threeparttable}
\usepackage{authblk}

\begin{document}

\maketitle

\begin{abstract}
\noindent
\textbf{Background}. In decision modelling with time to event data, there are a variety of parametric models which could be used to extrapolate the survivor function. Each of these implies a different hazard function and in situations where there is moderate censoring, they can result in quite different extrapolations. Expert opinion on the long-term survival or other quantities could reduce model uncertainty. \textbf{Objective}. We present a general and easily implementable approach for including a variety of types of expert opinions. \textbf{Methods}. Expert opinion is incorporated by penalizing the likelihood function. Inference is performed in a Bayesian framework, however, this approach can also be implemented using frequentist methods. The issue of aggregating pooling expert opinions is also considered and included in the analysis. \textbf{Results}. We validate the method against a previously published approach and include a worked example of this method. \textbf{Conclusions}. Expert opinions can be implemented in a straightforward manner using this approach, however, more work is required on the correct elicitation of these quantities.

\end{abstract}


\textbf{\textit{Keywords:}}
 Survival Models; Extrapolation; Expert Opinion.


\section*{Highlights}
\begin{itemize}
  \item Presentation of a novel and open-source method to incorporate expert opinion into decision modelling.
  \item Extends upon earlier work in that expert opinion can be incorporated with any parametric models.
  \item Provides methodological guidance for including expert opinion in decision modelling, which is a research focus area in NICE TSD 21 \cite{rutherford.2020}.
\end{itemize}

\section{Introduction}

The primary aim of many studies is to analyze the time until a pre-specified event of interest occurs. In these settings, the response variable is the time until that event, which is often called failure time, survival time, or event time. Time to event data are usually not observed for all observations under study, primarily  because the data from a study are analysed at a point in time when some individuals are still alive, resulting in these observations being censored.  

In clinical trials which have a time-to-event outcome, the primary objective is to identify if there is a statistically significant difference in the expected survival times of the treatment arm. In other disciplines such as health economics, the primary focus is to assess the long-term expected survival of both treatment groups so that the incremental health outcome of an intervention can be calculated. Except in situations where we are willing to assume that the long-term difference in health outcomes is similar to the differences in survival observed in the trial (see  \cite{Monnickendam.2019}), we are required to assume a parametric form for the data. These parametric models provide survival functions to predict the long-term survival and calculate the average time survived.    

Any valid probability distribution which has a support from $[0,\infty)$ can in principle be used for this purpose, and each distribution implies a particular functional form of the hazard function. From the exponential distribution, which assumes a constant hazard, to the four parameter generalized-F distribution which can accommodate bathtub type hazards, the choice of model will determine the hazard function and consequently the expected survival. Differences in long-term predictions can be particularly pronounced when a high proportion of the survival times are censored and may produce clinically implausible survival estimates. This issue has been discussed by \cite{Davies.2013} among others, and a number of solutions have been proposed, including model averaging \citep{Jackson.2010}, using external data \citep{Guyot.2017}, and expert opinion \citep{Cope.2019}.


In this paper we focus on how estimates of long-term survival of a treatment provided by clinicians can be incorporated with the parametric models fit to the trial data. While this topic has been considered by \cite{Cope.2019} the approach we describe does not require the discretization of the hazard function and is straightforward to apply to any parametric model. Furthermore, we discuss approaches to pool the elicited expert opinion which we assume can follow a wide variety of distributions rather than just the normal distribution considered by \cite{Cope.2019}.


Much of the initial work on this topic is from the area of reliability analysis, in which the predictions about the life of products or machinery in the population are computed by fitting a statistical distribution (often Weibull) to life data from a representative sample of units. \cite{Apostolakis.1979} calculated a rate for nuclear reactor melts by updating a gamma prior with the counts of accidents and near misses (a Poisson likelihood). \cite{Singpurewalla.1988} discuss a reparameterized Weibull model with the focus on estimating the median survival using a chi-squared prior. \cite{Campodonico.1993} estimate the median using a normal prior, while also eliciting opinion for the ``ageing'' parameter of the Weibull distribution. 

\cite{Coolen.1996} considers survival models for which conjugate priors exist (exponential, Gamma, and Weibull) although for two-parameter models, this requires assuming that one of these parameters is known. For the Weibull example, the author provides an example  where hypothetical assessments are made about survival times and an informative conjugate prior is derived from this information. 

\cite{Bousque.2006} estimates a Weibull distribution based on expert opinion on either the mean, mode and quantiles of the marginal distribution of time. By reparameterizing the Weibull distribution, the expert's implied opinion is scaled by an effective sample size which encodes the uncertainty of the opinion. This approach has the advantage that it does not require the expert to think about the mean and variance of the ``ageing'' parameter, however, it relies on the existence of conjugate distributions for survival models which are often not available in practice.


\cite{Ouwens.2018} incorporated expert opinion about survival probabilities at a particular timepoint for one and two parameter models by re-expressing one of the parameters in terms of the survival probability at the elicited timepoint and the other parameter (if applicable). The approach samples both a survival probability from the expert's prior distribution and the second parameter from its prior and uses these to calculate the first parameter. Although this can be implemented in programs such as Stan \citep{Stan.2020}, the approach is not possible if none of the parameters can be expressed in terms of the survival function and other parameters. This is the case for the Gamma distribution and any $\geq 3$ parameter models (i.e. Generalized Gamma, where parameters refer to the number of parameters in the probability distribution and not potential covariates on the location parameter). Although it is possible to use an optimizer to obtain a value of the first parameter conditional on the other parameters which will give the proposed survival, this requires writing custom Markov chain Monte Carlo (MCMC) algorithms. 

\cite{Cope.2019} introduced a method to incorporate expert information regarding survival probabilities when it has been provided at multiple timepoints. A Bayesian approach is used to fit a hazard function to the observed data and the hazards implied by the long-term survival beliefs of the expert. Weibull, Gompertz, and $1^{st}$ and $2^{nd}$ order polynomials can be fit with this approach using the JAGS statistical program \citep{Plummer.2003}. \cite{Cope.2019} re-express the individual survival times into multiple discrete time periods in which the hazard is assumed to be constant, however, if individual patient level data is available 
it is more efficient to analyse the data in its original format.


In this paper we consider how to incorporate multiple experts's opinions about survival probabilities at multiple timepoint(s) into a survival analysis. The approach is generalizable to any quantity for which an expert can provide an opinion e.g. median or mean survival times, that is consistent with the SHELF framework, and is compatible with a broad range of survival models. 
%
%
%
All methods outlined in this paper are available for use as an R package \citep{R} at \url{https://github.com/Philip-Cooney/expertsurv}.

The rest of the paper is organized as follows. We describe the required notation and discuss the proposed statistical method in Section \ref{stat-method}. In Section \ref{Multi-Experts} we discuss considerations when aggregating the opinions of multiple experts. Section \ref{Case-study} present an application of the statistical method to a real world example. Section \ref{Discussion} concludes the paper with a discussion of the key ideas in the papers along with summarizing some of the the challenges eliciting expert opinion. In the Appendix we validate our approach with a previously published example whereby expert opinion on median life was incorporated into the survival function.

\section{Survival Analysis with Expert Information}
\label{stat-method}



Suppose there are $n$ subjects under study, and that associated with each individual $i$ is a survival time $t_i$ and a fixed censoring time $c_i$. Each $t_i$ is assumed to be independent and identically distributed (i.i.d) with density $f(t_i)$ and survival function $S(t_i)$. The exact survival time of an individual will be observed only if $t_i \leq c_i$ and if not $t_i = c_i$ with the status indicated by  

\begin{equation*}
  \nu_i =
    \begin{cases}
      1 & \text{if } t_i \leq c_i\\
      0 & \text{if } t_i > c_i.
     \end{cases}       
\end{equation*}
For a parametric survival model, with associated set of parameters $\boldsymbol{\theta}= (\theta_1,\dots,\theta_p)'$ the likelihood function given the observed data $D$, is $ L(\boldsymbol {\theta}|D) = \prod_{i=1}^n f(t_i)^{\nu_i}S(t_i)^{1-\nu_i} = \prod_{i=1}^n h(t_i)^{\nu_i}S(t_i),$ where $h(t) = \frac{f(t)}{S(t)}$ denotes the hazard function. In a Bayesian analysis we assume a prior distribution for $\boldsymbol {\theta}$ denoted by $\pi(\boldsymbol {\theta})$. The posterior distribution is then $\pi(\boldsymbol{\theta}|D) \propto L(\boldsymbol{\theta}|D)\pi(\boldsymbol{\theta}).$ 

As a running example, consider an exponential distribution, with associated hazard $h(t) = \theta$ and survival function $S(t) =  \exp\lbrace -\theta t \rbrace$. The likelihood of an exponential model is then $L( {\theta}|D) = \theta^{\sum_{i = 1}^n \nu_i} \exp \left \lbrace-\theta \sum t_i\right \rbrace.$ If the prior distribution for $\theta$ has been specified as $\mathcal{G}(\alpha, \beta),$ i.e., a Gamma density with parameters $\alpha$ and $\beta$, then the posterior distribution is available in closed form as Gamma distribution $\mathcal{G}(\alpha +\sum_{i = 1}^n \nu_i,\beta +\sum_{i=1}^n t_i)$. While in this case the posterior distribution is tractable, Bayesian inference for other distributions is more challenging and relies on modern computational methods for inference. Even in this case, tractable inference requires specification of the prior in a specific framework that will not be intuitive to a non-specialist. See Table~\ref{tab:my_label} for a full list of the survival models under consideration in this paper.

\subsection{Integrating Expert Opinion with Trial Data} \label{expert-opinion}


\noindent
Consider the situation where an expert has an opinion about the survival probability at a particular time $t^*$. This is in contrast to the previous example in Section~\ref{Multi-Experts} where expert opinion was elicited explicitly in the form of a (conjugate) prior distribution. We propose to incorporate this information into the analysis by recasting the prior distribution as an explicit function of the timepoint $t^*$, and implicitly as a function of the parameters of interest $\boldsymbol{\theta}$.


To fix this idea, consider an exponential model being fit to data,  with a normal distribution with mean $\mu_{\text{expert}}$ and variance $\sigma_{\text{expert}}^2$ describing the expert's belief about survival at a particular timepoint $t^*$, so that $S(t^*)
\sim\mathcal N(\mu_{\text{expert}}, \sigma_{\text{expert}}^2)$. The posterior density is then proportional to:
$$\pi( {\theta}|D, \mu_{\text{expert}},\sigma_{\text{expert}}^2) \propto L( {\theta}|D)\pi_{t^*}(\theta | \mu_{\text{expert}},\sigma_{\text{expert}}^2)
\pi(\theta),$$
\noindent
where $$\pi_{t^*}(\theta | \mu_{\text{expert}},\sigma_{\text{expert}}^2)$$ denotes the expert opinion regarding survival at time $t*$ and $\pi(\theta)$ denotes a more standard, typically weakly informative, prior for $\theta$. For an exponential model the survival at the elicited timepoint is $S(t^*) = \exp\lbrace -\theta t^*\rbrace$ so that
$$\pi_{t^*}(\theta | \mu_{\text{expert}},\sigma_{\text{expert}}^2) \propto \exp \left \lbrace -\frac{1}{2}\Bigg(\frac{\exp(-\theta t^*) - \mu_{\text{expert}}}{\sigma_{\text{expert}}}\Bigg)^2\right\rbrace.$$ While the resultant posterior does not have closed form, this is not of practical importance when using modern computational Bayesian methods. More generally, the advantage of this approach is that it can be applied using a wide family of survival models, including those with 3 or more parameters. It is also straightforward to use different types of elicited probability distribution,   as well as incorporate additional timepoints. The contribution of the expert opinion can be implemented in programmes like WinBUGS and JAGS (through the zeros trick) or Stan (by simply incrementing the log probability). \footnote{We were not able to implement the Gamma, Gompertz or Generalized Gamma models with expert opinion on the survival function or expected survival in Stan  and we therefore implement these models in JAGS (see Appendix for further details).}

Although we have presented this method in a Bayesian framework, it can also be motivated from a frequentist perspective as an example of a penalized likelihood method \citep{Cole.2013}. In this framework, we impose additional constraints on the parameter space by modifying the likelihood so that it is a function of the observed data and of pseudo-observations based on the elicited predicted survival at the timepoint $t^*$. We can estimate model parameters using standard optimisation techniques \citep{nocedal2006}.

\subsection{Incorporating Multiple Expert Opinions}
\label{Multi-Experts}

In many situations it is advantageous to consider opinions from multiple experts. There are different approaches to include this information as a prior within a Bayesian framework. \cite{OHagan.2006} discuss both this topic and a general framework for eliciting probabilities. One such technique is opinion pooling, where a $\emph{consensus distribution }$ for $\pi(\theta)$ is obtained as some function of the distributions ${\pi_1(\theta),\dots,\pi_m(\theta)}$ elicited from each of the $m$ individual experts. The consensus distribution is then used as the prior for the analysis. 

A logarithmic opinion pool is obtained by taking a weighted $\emph{geometric}$ mean of the distributions, 
$$\pi(\theta) \propto \prod_{j=1}^m \pi_j(\theta)^{w_j},$$ with weights specified such that $\sum^m_{j=1} w_j = 1.$ When the decision maker is equally confident in the abilities of all experts, it is common to choose $w_j = 1/m$ for all $j$. The advantage of this approach is that it is $\emph{externally Bayesian}$. When new data are obtained, one could either update each expert's distribution individually and then combine the resulting posterior distributions using logarithmic pooling, or first combine the expert's distributions and then update the consensus distribution. These will result in the same posterior distributions.

Continuing our example and assuming an exponential distribution with constant hazard, if $m$ experts have expressed their prior beliefs about $\theta$ as Gamma priors $\mathcal G(\alpha_j, \beta_j)$, $j=1, \ldots, m$ the pooled prior is also a Gamma distribution, $\mathcal{G}(\sum_{j=1}^k w_j \alpha_j,\sum_{j=1}^m w_j \beta_j),$ and the resulting posterior distribution is then $\mathcal{G}(\sum_{j=1}^m w_j \alpha_j + \sum_{i = 1}^n \nu_i,\sum_{j=1}^m w_j \beta_j +\sum_{i=1}^n t_i)$.  If we were to compute the posterior distribution using each expert prior separately, and then compute the logarithmic opinion pool post-hoc, it is evident that the same posterior distribution would be obtained. 



Another form of expert pooling is the $\emph{linear opinion pool}$
 $$\pi(\theta) = \sum_{j=1}^m w_j \pi_j(\theta),$$
 \noindent
 which is the weighted arithmetic mean of the distributions. This approach is not externally Bayesian. Continuing our example, a weighted sum of Gamma distributions is not a Gamma distribution and is not available in an analytic form unless the rate parameters are equal \citep{DiSalvo.2008}. 
 
\begin{figure*}[tbp]
    \centering
    \begin{subfigure}[t]{0.5\textwidth}
        \centering
        \includegraphics[height=2in]{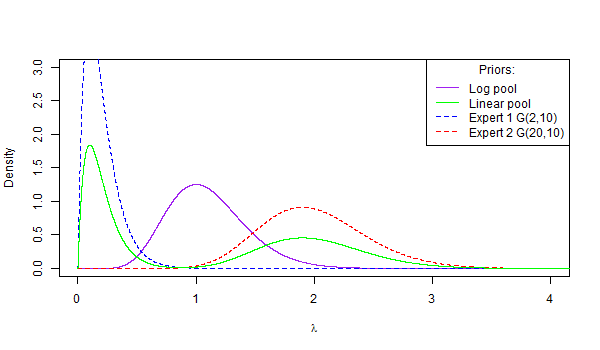}
        \caption{Linear and logarithm pooling of opinions}
        \label{fig:expert-prior}
    \end{subfigure}%
    ~ 
    \begin{subfigure}[t]{0.5\textwidth}
        \centering
        \includegraphics[height=2in]{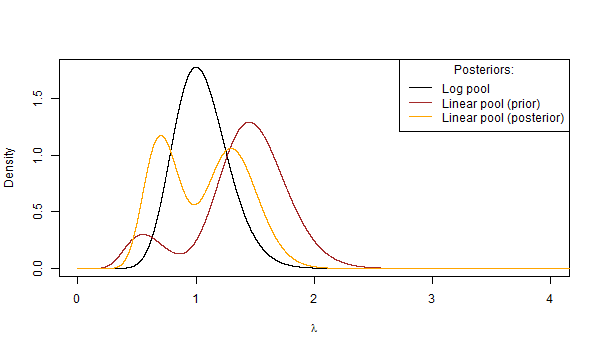}
        \caption{Posterior distributions for linear and logarithm pooling}
        \label{fig:expert-posterior}
    \end{subfigure}
    \caption{Aggregation of Expert Opinions}
     \label{fig:expert}
\end{figure*}

 Consider a hypothetical example where two experts have provided their opinions on $\theta$ for an exponential model, with the experts holding somewhat conflicting opinions. We suppose Expert 1 has a prior of $\mathcal{G}(2,10)$ and Expert 2 has a prior of $\mathcal{G}(20,10)$. Figure \ref{fig:expert-prior} presents both pooling approaches for the prior expert opinions while Figure \ref{fig:expert-posterior} shows the resulting posterior distributions. The linear pool results in two separate posterior distributions depending on whether pooling was conducted on the individual priors or individual posteriors. 
 
  It should be noted that logarithm pooling is associated with some disadvantages \citep[see][]{Williams.2021}, and in general linear pooling is more commonly used. \cite{OHagan.2006} notes that when using logarithmic pooling, the decision maker regards as implausible any values of $\theta$ that are considered implausible by any single expert. The linear opinion pool, on the other hand, concentrates more in the area where the experts opinions overlap, but it does not rule out values of $\theta$ that are supported by only one expert. These properties are evident in both plots in Figure~\ref{fig:expert}. 
 
 More generally, when using our approach as described in Section~\ref{expert-opinion}, when multiple expert opinions are available, the posterior has the form \[\pi( {\theta}|D) \propto L( {\theta}|D)\pi_{t^*}(\theta)\pi(\theta),\] where $\pi_{t^*}(\theta)$ denotes our pooled expert prior regarding survival at time point(s) $t^*$. If we use a linear method to pool our prior information, then the resultant posterior will be different then if we ran separate analyses using each expert opinion and pooled the results \emph{a posteriori}.

\section{Case Study: ELIANA trial}
\label{Case-study}

We apply the approach described in Section~\ref{stat-method} to aggregate experts opinions about survival at multiple timepoints and combine this information with the trial data to inform the long-term survival extrapolations. Both the data and elicited expert opinions have been previously presented by \cite{Cope.2019} and the focus of this analysis is to present the results using six commonly used parametric survival models along with the flexible Royston-Parmar spline models. We also address how the expert opinions can be linearly pooled and how this somewhat complex prior belief can be incorporated into the analysis. 

One point to note is that in the original publication \cite{Cope.2019}, expert opinion was elicited on the expected survival probabilities at 2, 3, 4 and 5 years for patients treated with tisagenlecleucel based on the available 1.5 year results from the ELIANA trial \cite{Grupp.2016}. In this analysis we consider the longer term follow-up from ELIANA based on median duration of follow-up of 24.2 months (range; 4.5-35.1 months) \cite{Grupp.2018}, and combine with the experts opinions on expected survival for years 4 and 5 (as we have an estimate of the survival function for times < 2.8 years). It is highly probable that the experts would calibrate their opinions on expected survival based on the longer term follow-up, however, for the purpose of this analysis we assume that these opinions are still valid.   

\begin{figure}[h!]
 \centering
 \includegraphics[scale=.7]{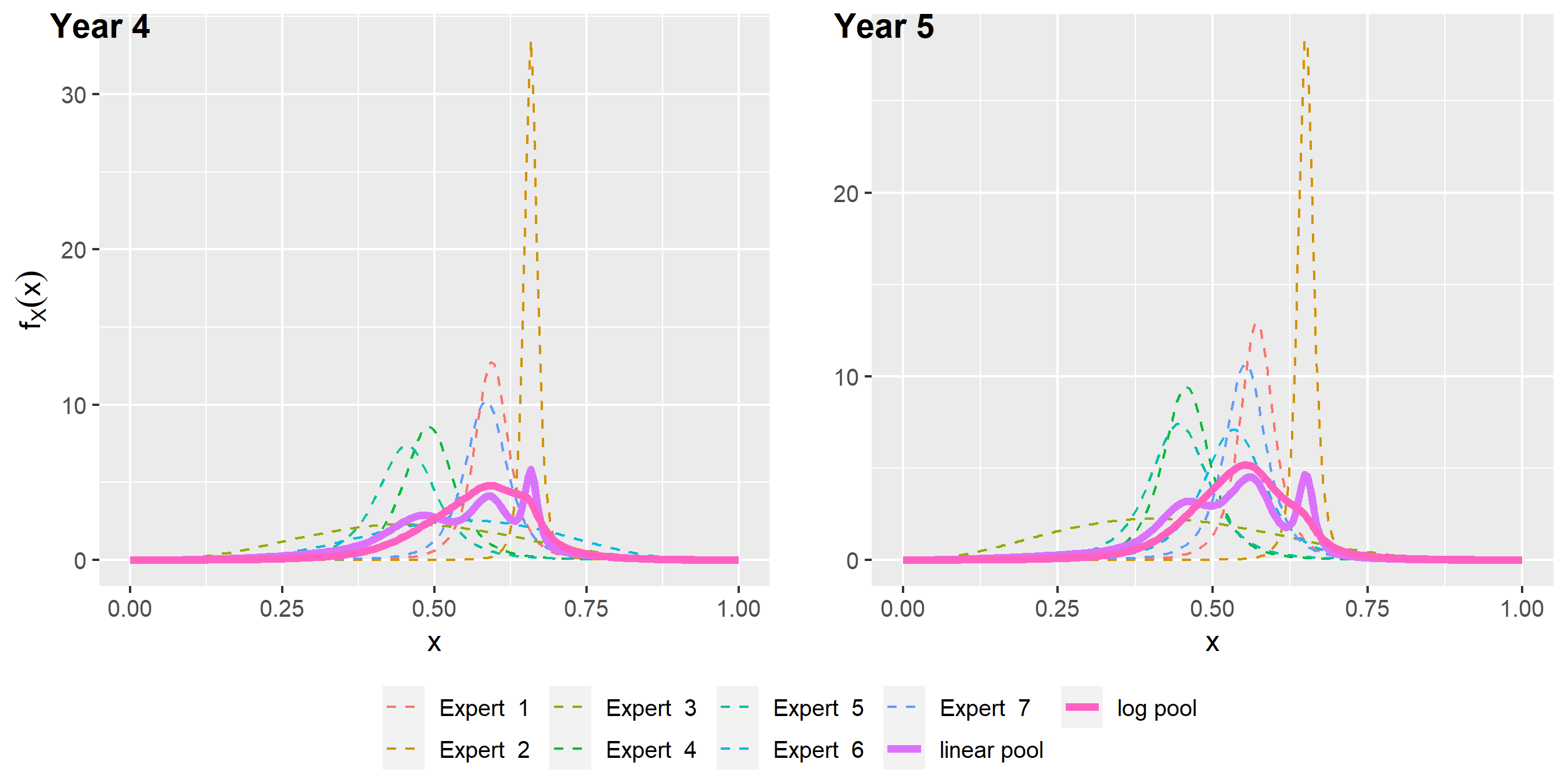}
 \caption{Best fitting distributions for Expert's Opinions and aggregated distributions}
 \label{fig:expert-opinion-Year4-5}
 \end{figure}

We first consider the expert beliefs for survival at each timepoint and identifying which distribution most accurately describes their elicited belief. In \cite{Cope.2019}, for each timepoint, experts were asked to first estimate the upper plausible limit (UPL), followed by the lower plausible limit (LPL) so that they are $99\%$ sure that the true survival probability is contained within that interval. Experts were also asked to estimate the most likely values (MLV). Rather than assuming that these beliefs could be adequately modelled with a normal distribution, we used the functions from the \emph{SHELF} package \cite{SHELF} to identify the best fitting distribution by minimizing the least square error. Because we wished to include the expert's MLVs, we modified these functions assuming that the MLV represented the mode of the distribution and included this quantity in the least squares optimization. The best fitting distribution was the one which minimized the least squares from either the normal, $t$, lognormal, Gamma or Beta distributions. The individual distributions for years 4 and 5 along with the logarithm and linear pooling are presented in Figure \ref{fig:expert-opinion-Year4-5}. The majority of the expert beliefs were described by t distributions with 3 degrees of freedom, with the remainder being described as beta distributions. 

Because linear pooling is the more common pooling method we will use this approach to combine the experts opinions. We see in Figure \ref{fig:expert-opinion-Year4-5} the variety of expert opinions, with Expert 1 and 7 similar (but Expert 7 has more uncertainty), while Expert 4 and 5 are also similar. Expert 6 is noteworthy as they have what appears to be a much more diffuse belief in Year 4 than Year 5. The reason for this is that their best fitting distribution in Year 4 and 5 are the beta and t distributions respectively. The UPL and LPL refer to $99.5\%$ and $0.5\%$ percentiles and the low degrees of freedom produces a heavy tailed t-distribution which can assign small probability to much of the parameter space while still appearing relatively concentrated. The beta distribution, however, must be much more diffuse to assign probability to the UPL and LPL. In this case it might be appropriate to pick one family of distributions for each expert, the distribution which minimizes total error across the timepoints. Across both timepoints Expert 2 has a very strong opinion. These collection of opinions result in tri-modal distribution for the linear pool. The logarithm pool is smoother and assigns lower probability at the more extreme ends of the parameter space.  


 
\begin{figure}[h!]
 \centering
 \includegraphics[scale=0.65]{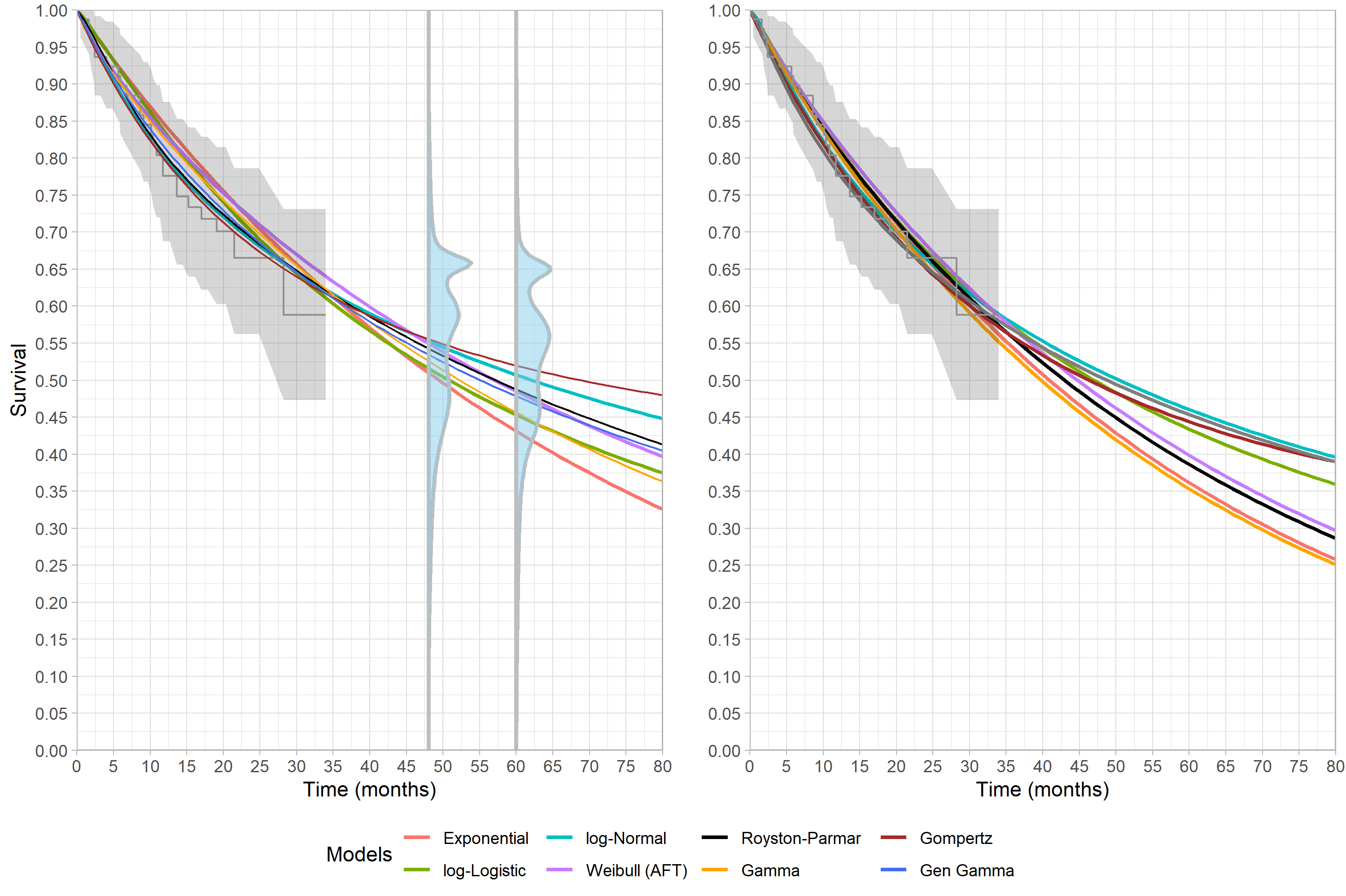}
 \caption{Right: Updated ELIANA trial data and expert opinion with extrapolated survival functions with pooled Prior beliefs shown at year 4 and 5. Left: Survival functions fit by maximum likelihood}
 \label{fig:surv-ELIANA}
 \end{figure}

 \noindent
 The survival models presented in Figure \ref{fig:surv-ELIANA} were fit to the data and expert opinion. For the models fit with Stan, inference was based on 3 chains each containing 10,000 iterations with the first 5,000 as burn-in, while for models fit with JAGS each chain contained 50,000 iterations and the first 10,000 discarded as burn-in. Models were also fit using maximum likelihood to highlight the effect of the expert opinions. As shown in Table \ref{tab:Mod-fit-ELIANA} the Log-Normal model has the best statistical fit with respect to the Deviance information criterion (DIC) \citep{Spiegelhalter.2002}).  Models which allowed for rapidly decreasing hazards (Gompertz) or non-monotonic hazards (e.g. Log-Normal or Log-Logistic) seem to provide the best fit to the experts opinions and the data, a property which all of the four best fitting models have. Within these four models the differences in DIC is $<3$, suggesting that they are broadly similar in model fit. Considering the models estimated via maximum likelihood, the exponential model which assumes a constant hazard was the best fit according to Bayesian Information Criteria (BIC). Including the experts opinions assigns substantial probability to a high long-term survival probability and the parametric models which accommodate lower longer term hazards fit the data and expert opinions the best.

\begin{table}[h!]

\caption{Survival models ordered by DIC (lower is better)}
\centering
\begin{tabular}[t]{l|r|r}
\hline
Models & DIC & BIC\\
\hline
Log-Normal & 272.09 & 282.91\\
\hline
Gompertz & 273.69 & 283.28\\
\hline
Log-Logistic & 274.07 & 283.20\\
\hline
GenGamma & 274.22 & 287.25\\
\hline
Exponential & 275.14 & 279.55\\
\hline
Royston-Parmar (1-knot) & 275.73 & 283.83\\
\hline
Gamma & 276.55 & 283.88\\
\hline
Weibull ATF & 278.72 & 283.83\\
\hline
\end{tabular}
\label{tab:Mod-fit-ELIANA}
\end{table}

\section{Discussion}
\label{Discussion}
This paper extends previous work by \cite{Cope.2019} and \cite{Ouwens.2018} to a full range of parametric models. We also provide full code to implement the method. Using our approach it is straightforward to incorporate information about other quantities of interest (e.g. median, mean survival or mean survival difference) into an analysis. Additionally this paper highlights important considerations with respect to pooling information from multiple experts and how multi-modal aggregated distributions can be incorporated in the analysis.

The analysis presented in Section \ref{Case-study} results in similar conclusions as the analysis performed by \cite{Cope.2019}. In their analysis the preferred model was the Gompertz model which is also the second ranked model in our analysis (based on DIC), with the Log-Normal ranked highest. In the approach presented by \cite{Cope.2019}, estimates from each expert were modelled separately, and the overall estimate reflected a combined overall distribution. This necessitated fitting models for each of the individual experts before combining the results. The authors noted that this approach avoids pooling or a model averaging, which would provide narrower intervals around the mean. We would argue that their approach does not avoid pooling, rather it is a linear pool of the posterior distributions and as our illustrative example in Figure \ref{fig:expert-posterior} shows does not automatically lead to narrower intervals. Furthermore, we believe a decision maker would value an aggregated prior as it would aid understanding about how this prior changes the analysis compared an analysis using the data alone.

As noted by \cite{OHagan.2019} we generally prefer the outcome that we are eliciting to be a single probability distribution representing the combined knowledge of experts in the field. Resolving the experts’ judgments into a single distribution is known as the problem of aggregation. In this paper we use mathematical aggregation (as we do not have access to the experts), however, we note that the SHELF framework uses behavioural aggregation in which the group of experts to discuss their knowledge and opinions, and to make group ``consensus'' judgments, to which an aggregate distribution is agreed. Even in situations where behavioural aggregation is the objective, using a mathematical aggregation of the experts opinion might be a useful visual tool in agreeing the consensus distribution.

Although expert opinion can be of value in reducing the differences in extrapolated survival probabilities for different parametric models, the appropriate elicitation of these quantities is challenging. One important point is how the questions are framed, with \cite{Bousque.2006} providing examples of some open questions which are relevant when eliciting beliefs about survival. Clearly defined elicitation questions are particularly relevant as the experts may not be familiar with statistical terms. For example, \cite{Bousque.2006} notes that experts often misinterpret averages as medians.  It has also been frequently discussed that experts can be overconfident \citep{OHagan.2019,Bousque.2006,Lin.2008}, with \cite{Lin.2008} suggesting that calibration and differential weighting of experts may reduce this overconfidence. Within this analysis it is possible that Expert 2 provided survival estimates that were overconfident, and exclusion of this experts opinion slightly lowers the expected survival estimates, although the ordering of DIC for the parametric models remains the same. 

Discussing this point further, we may be able to address the issue of overconfidence if we can provide the the expert with feedback on the information contained within the prior. An intuitive measure of the information is the effective sample size (ESS) which is the sample size required to obtain parameters (when the priors are vague) with the same standard deviation as the elicited priors. The area expert may use this as a basis to modify his/her judgments, if desired. In the context of our method, the most straightforward approach is to note that that a $\mathcal{B}(\alpha, \beta)$ (beta) distribution can naturally be used to represent survival probabilities at a specific time point, and that such distributions have an effective sample size (ESS) of $\alpha + \beta$. We can interpret the ESS as the sample size of patients whose survival times we are able to  observe (without censoring) until the elicited timepoint (with those surviving after that time considered censored). This ESS will incorporate the information the expert obtained having seen the short term observed data plus their opinions on the long-term survival. If we look at the ESS of the beta distributions that were fit to each of the experts survival opinions at year 5 we see that experts 3 to 7 had ESS between 8 and 61. Expert 2 had an ESS of 263, suggestive of overconfidence given that the sample size of the trial is 75 and therefore may need to be re-calibrated.  

Although not discussed in this paper, there are situations where the expert may have considerable experience with the comparator arm and may be more comfortable providing an opinion on the plausible survival probabilities for the comparator at particular timepoint(s). If a relationship such as proportional hazards (PH) or accelerated time factor (ATF) can be considered tenable (i.e. evaluated based on trial data and assume to hold in the long-term), a survival model with the PH or ATF parameterization with treatment status as a covariate could be estimated. Alternatively experts may be willing to provide an estimate of the expected survival difference between two treatments. Both of these approaches are possible with the approach and we provide simulated examples of each situation with the code. 

 It is clear that further research on elicitation of long-term survival probabilities and best practices are important if expert opinions are to become more widely used in health technology assessments using time to event outcomes. Further research on deriving ESS of priors for long-term survival probabilities and other quantities which may be elicited would also aid the accurate elicitation of expert beliefs. As noted in Section \ref{stat-method}, similar results can be obtained for frequentist methods and as next steps we hope to modify the \emph{flexsurv} package to facilitate this.       


\section*{Software}

All methods outlined in this paper are available for use at  \url{https://github.com/Philip-Cooney/expertsurv}. The code in the package is adapted from \texttt{survHE} \citep{suvrHE.2020} which primarily considers survival analysis from a Bayesian approach. we allow a user to supply opinions on survival timepoints and expected differences . The package is a wrapper for the \texttt{flexsurv} package \citep{flexsurv.2016}, meaning that the user can easily see the influence that priors have on the analysis and more generally making the package straightforward to use.

\appendix
\section*{Appendix}

\subsection*{Implementation of distributions in JAGS}


We could not fit several of the models in Stan and therefore we fit the Gamma, Gompertz and Generalized Gamma distributions in JAGS. Below we describe how we analytically evaluate the expected survival for the Gompertz and Generalized Gamma distributions.

We see the expectation of the Gompertz distribution is given by \cite{siegrist} as:
$b\exp{a}E_a(-1)$ where $E_a(t) = \int_1^\infty u^t\exp(-au) du$. Note that $a,b$ are different from the flexsurv parameterization (denoted as  $a^*,b^*$) such that $a = \frac{b^*}{a^*}$ and $b = 1/a^*$.
\noindent
We note that $E_a(t) = \Gamma(0,a)$ (upper incomplete Gamma function) \cite{wolfram.mathworld}. We need to approximate this function, and note that by definition $\Gamma(x) = \Gamma(x,a) + \gamma(x,a)$. Hence, $\Gamma(0,a) = \lim_{x \to 0}(\Gamma(s) - \gamma(s,a))$. We can compute the incomplete gamma function as $\Gamma(s)\text{CDF}_{\Gamma}(X = a,\alpha = s,\beta = 1)$.
\noindent
We have the expected survival as $(1/a^*)\exp{b^*/a^*}\Gamma(0,b^*/a^*)$.

For Generalized Gamma the parameterization in JAGS is slightly different to \cite{Stacy.1962} with $b \times r = d$, $b = p$ and $\lambda = 1/a$ who gives the the mean as $\frac{\Gamma((b\times r)/b)}{\lambda \gamma(r)}$. For consistency of results the \texttt{flexsurv} package we have $\mu = -\log(\lambda) +\log(r)/b$, $\sigma = 1/(b*\sqrt{r})$ and $Q =\sqrt{1/r}$.

\subsection*{Validation of approach}

In this example we compare the results of \cite{Singpurewalla.1988} and the method we discuss above. In \cite{Singpurewalla.1988} they consider a Weibull distribution with a proportional hazards (PH) parametrization (see Appendix for details). The median survival time for this distribution is $t_{0.5} =  \frac{-ln(0.5)}{m}^{\frac{1}{a}}$ which for clarity we will denote as $\kappa$. If we re-express the distribution in terms of $\kappa$ we obtain a survival function $S(t) = \exp{\big\{ln(0.5)\big(\frac{t}{\kappa}^a\big)\big\}}$ and hazard function $h(t) =\frac{-ln(0.5)at^{a-1}}{\kappa^a}$. From this we can obtain the likelihood of this data using the expressions in Section \ref{expert-opinion}.

\begin{figure}[h!]
\centering
\includegraphics[scale=0.7]{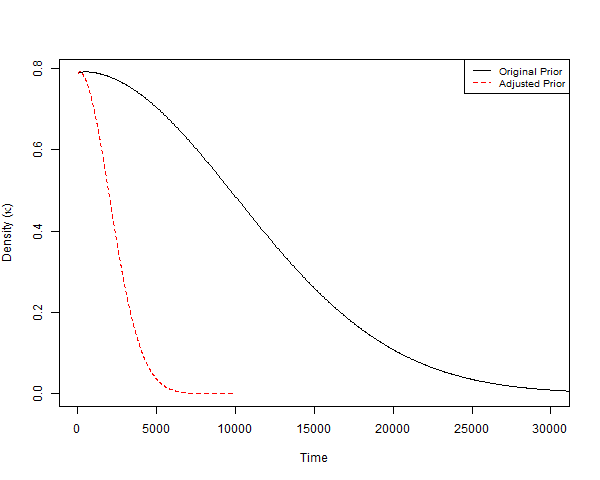}
\caption{Original prior used by (Singpurewalla et. al 1988) and adjusted prior}
\label{fig:prior-example}
\end{figure}

\begin{figure*}[h!]
    \centering
    \begin{subfigure}[t]{0.5\textwidth}
        \centering
        \includegraphics[height=3.3in, width = 3.3 in]{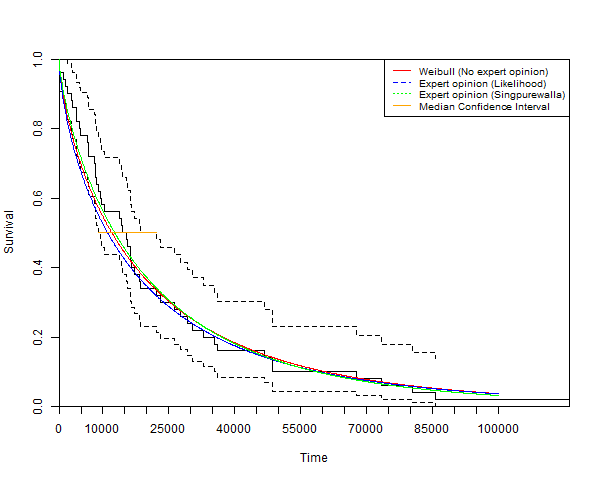}
        \caption{Survival functions under the original prior}
        \label{fig:posterior-example1}
    \end{subfigure}%
    ~ 
    \begin{subfigure}[t]{0.5\textwidth}
        \centering
        \includegraphics[height=3.3in, width = 3.3 in]{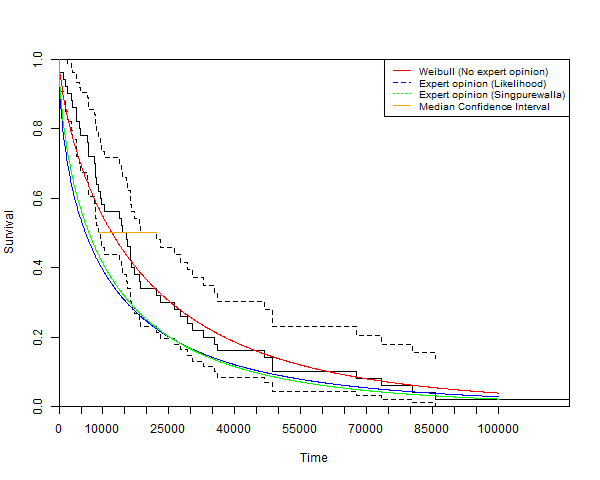}
        \caption{Survival functions under the adjusted prior}
        \label{fig:posterior-example2}
    \end{subfigure}
    \caption{Comparison of Approaching incorporating expert opinion}
    \label{fig:posterior-examples}
\end{figure*}

The experts belief about $\kappa$ is characterized by the location or mean $l$ and standard deviation $s$.\cite{Singpurewalla.1988} also consider to additional parameters $c, v$ which can be used to calibrate the expert's opinion about $l$ and $s$, however, in the case the analyst does not wish to modulate the expert's opinion then $c = 1$ and $v = \frac{1}{2}$. Under some mild assumptions $$[c^2v/(sl^2)]\kappa^2 \approx \chi^2((v/s)+1)$$ and assuming no calibration of the expert's opinion and noting that the square root of a $\chi^2(n)$ random variable is a $\chi(n)$ random variable $$\big[1/(\sqrt{(2s)}l)\big]\kappa \approx \chi((v/s)+1).$$
\noindent
Therefore $\kappa$ is a $\chi((v/s)+1)$ random variable scaled by $\sqrt{(2s)}l$ and we assume that $a \sim \mathcal{G}(\alpha, \beta)$ both parameters are specified. \cite{Singpurewalla.1988} describe a Bayes estimator for the parameters using some approximations, however, it is straightforward to use JAGS/Stan to obtain the complete posterior distribution. Using simulated data they provide in the paper, they set $l = 500$ and $s=200$ and did not assume any modulation of the expert's opinion. This gives a prior (termed Original Prior) in Figure \ref{fig:prior-example} and posterior survival curves in Figure \ref{fig:posterior-example1}. The fact that the posterior distributions are very similar to the analysis without any expert opinion is unsurprising as the original prior had a significant probability within the $95\%$ confidence interval implied by the data alone (8909 - 22188). In a second example we adjust the prior belief of the expert to yield a much lower median value (termed adjusted prior in Figure \ref{fig:prior-example}) and see that the mean survival posterior for both approaches incorporating expert opinion are very similar and as expected, outside the confidence interval for the median (Figure \ref{fig:posterior-example2}).

As can be seen from this example, specifying a prior belief using a $\chi$ distribution is not particularly intuitive, however, it highlights that the proposed approach is consistent with previous methods and can be easily adjusted to include prior beliefs (using any distribution) on any quantity of interest.   
\newpage
    \clearpage
    \thispagestyle{empty}
    \newgeometry{vmargin={0.5cm}, hmargin = {0.5cm}}
    \begin{landscape}
     \begin{threeparttable}
     \begin{center}
    \centering
    
      \caption{Flexsurv Parametrization of Survival Models}
     \label{tab:my_label}
    \begin{tabular}{|c|c|c|c|c|c|}
    \hline
 &\thead{PDF} & \thead{CDF} & \thead{Hazard} & \thead{Parameters} & \thead{Hazard shape} \\	
   \hline
 Exponential   & $\lambda e^{-\lambda t}$ & $1 - e^{-\lambda t}$ & $\lambda$ & $\textcolor{red}{rate}  = \lambda  > 0$ & Constant  \\
      Weibull (AFT) & $\frac{a}{b}\left(\frac{t}{b}\right)^{a-1}e^{-(t/b)^a}$ & $1 - e^{-(t/b)^a}$&1 $\frac{a}{b}\left(\frac{t}{b}\right)^{a-1}$& \makecell{$\text{shape} = a  > 0$ \\  $\textcolor{red}{scale} = b > 0 $}& \makecell{Constant, \\monotonically increasing/decreasing} \\
     $ \text{Weibull (PH)}^2$  & $a m t^{a-1} e^{-m t^a}$ & $1 - e^{-mt^a}$& $amt^{a-1}$ & \makecell{$\text{shape} = a  > 0 $ \\  $\textcolor{red}{scale} = m > 0$} &  \makecell{Constant, \\monotonically increasing/decreasing} \\
      Gompertz      & $b e^{at} \exp\left[-\frac{b}{a}(e^{at}-1)\right]$      & $1 - \exp\left[-\frac{b}{a}(e^{at}-1)\right]$ & $b e^{at}$ & \makecell{$\text{shape} = a \in (-\infty, \infty)$ \\  $\textcolor{red}{rate} = b > 0$} & \makecell{Constant, \\monotonically increasing/decreasing} \\
       $\text{Gamma}^3$         & $\frac{b^a}{\Gamma(a)}t^{a -1}e^{-bt}$& $\frac{\gamma(a, bt)}{\Gamma(a)}$ & f(t)/S(t)& \makecell{$\text{shape} = a > 0 $ \\  $\textcolor{red}{rate} = b > 0$} & \makecell{Constant, \\monotonically increasing/decreasing} \\
       Lognormal     & $\frac{1}{t\sigma\sqrt{2\pi}}e^{-\frac{(\ln t - \mu)^2}{2\sigma^2}}$ & $\Phi\left(\frac{\ln t - \mu}{\sigma}\right)$& f(t)/S(t) & \makecell{ $\textcolor{red}{meanlog} = \mu \in (-\infty, \infty)$ \\   $\text{sdlog} = \sigma > 0$ }& \makecell{Arc-shaped,\\ monotonically decreasing}\\
       LogLogistic   & $\frac{(a/b)(t/b)^{a-1}}{\left(1 + (t/b)^a\right)^2}$ & $\frac{1}{(1+(t/b)^a)}$ & $1-\frac{(a/b)(t/b)^{a-1}}{\left(1 + (t/b)^a\right)}$& \makecell{ $\text{shape} = a > 0$ \\  $\textcolor{red}{scale} = b > 0$} &
            \makecell{Arc-shaped,\\ monotonically decreasing}\\
      Generalized $\text{Gamma}^{3,4}$ & $\frac{|Q|(Q^{-2})^{Q^{-2}}}{\sigma t \Gamma(Q^{-2})}  \exp\left[Q^{-2}\left(Qw-e^{Qw}\right)\right] $&
             \makecell{$\frac{\gamma(Q^{-2}, u)}{\Gamma(Q^{-2})} \text{ if } Q \neq 0 $ \\ 
              $\Phi(w) \text{ if } Q = 0 $} & f(t)/S(t) &
              \makecell{$\textcolor{red}{mu} = \mu \in (-\infty, \infty)$ \\ $\text{sigma} = \sigma > 0 $ \\ $\text{Q} = Q \in (-\infty, \infty)$} &
              \makecell{Arc-shaped,\\ bathtub-shaped,\\ monotonically increasing/decreasing} \\
      Generalized $\text{F}^5$ & $\frac{\delta(m_1/m_2)^{m_1}e^{m_1 w}}{\sigma t (1 +m_1e^w/m_2)^{m1+m2}B(m_1,m_2)}$ & $\int_0^{m_2(m_2+m_1e^w)^{-1}}\frac{t^{m_2-1}(1-t)^{m_1 -1}}{B(m_2,m_1)}dt$ & f(t)/S(t) & \makecell{$P \ge 0$,\\ Others as per Gen Gamma} & \makecell{Arc-shaped,\\ bathtub-shaped,\\ monotonically increasing/decreasing} \\
              
     Royston-Parmar Splines  &  & &  & See \cite{Royston.2002} & Almost any shape \\           
     \hline         
     \end{tabular}
     \begin{tablenotes}
     \item[1]  Red colour refers to location parameter.
     \item[2] The proportional hazard (PH) model is a reparameterization of the accelerated failure time (AFT) model with $m = b^{-a}$.
     \item[3] $\Gamma(z) = \int_{0}^{\infty} x^{z-1}e^{-x} dx $ is the gamma function, $\gamma(s,x) = \int_{0}^{x} t^{s-1}e^{-t}dt$ is the lower incomplete gamma function.
     \item[4] $w = (log(t) - \mu)/\sigma, u = Q^{-2}e^{Qw}$ and $\Phi$ is the cumulative normal distribution function.
     \item[5] $w = (log(t) - \mu)/\sigma,\delta = (q^2 + 2p)^{1/2}, m_1 = 2(q^2 +2p +q\delta)^{-1}, m_2 = 2(q^2 +2p - q\delta)^{-1}$ and $B()$ is the beta function. 
        \end{tablenotes}
        \end{center}

        \end{threeparttable}
    \end{landscape}
    \clearpage
    \restoregeometry


\end{document}